\documentclass[aps,pof,superscriptaddress]{revtex4}
\usepackage[parfill]{parskip}    			
\usepackage{amssymb}
\usepackage{amsmath}
\usepackage{graphicx}					
\usepackage{epstopdf}					
\usepackage[latin1]{inputenc}				
\usepackage[T1]{fontenc}					

\usepackage{version}				
\usepackage{cases}						

\usepackage{pstricks}					

\usepackage{mathrsfs}
\usepackage[squaren, Gray]{SIunits}			

\begin{document}

\title{New criteria to detect singularities in experimental incompressible flows}
\author{Denis Kuzzay}
\affiliation{
SPEC/IRAMIS/DSM, CEA, CNRS, University Paris-Saclay, CEA Saclay, 91191 Gif-sur-Yvette, France
}
\author{Ewe-Wei Saw}
\email{ewe-wei.saw@cea.fr}
\affiliation{
SPEC/IRAMIS/DSM, CEA, CNRS, University Paris-Saclay, CEA Saclay, 91191 Gif-sur-Yvette, France
}
\author{Fabio J. W. A. Martins}
\affiliation{
Laboratoire de Mécanique de Lille, France
}
\author{Davide Faranda}%
\affiliation{
Laboratoire des Sciences du Climat et de l'Environnement, LSCE/IPSL, CEA-CNRS-UVSQ, Université Paris-Saclay, F-91191 Gif-sur-Yvette, France
}
\author{Jean-Marc Foucaut}
\affiliation{
Laboratoire de Mécanique de Lille, France
}
\author{François Daviaud}%
\affiliation{
SPEC/IRAMIS/DSM, CEA, CNRS, University Paris-Saclay, CEA Saclay, 91191 Gif-sur-Yvette, France
}

\author{Bérengère Dubrulle}%
\affiliation{
SPEC/IRAMIS/DSM, CEA, CNRS, University Paris-Saclay, CEA Saclay, 91191 Gif-sur-Yvette, France
}

\begin{abstract}

We introduce two new singularity detection criteria based on the work of Duchon-Robert (DR) [J. Duchon and R. Robert, Nonlinearity, 13, 249 (2000)], and Eyink [G.L. Eyink, Phys. Rev. E, 74 (2006)] which allow for the local detection of singularities with scaling exponent $h\leqslant1/2$ in experimental flows, using PIV measurements. We show that in order to detect such singularities, one does not need to have access to the whole velocity field inside a volume but can instead look for them from stereoscopic particle image velocimetry (SPIV) data on a plane.  We discuss the link with the Beale-Kato-Majda (BKM) [J.T. Beale, T. Kato, A. Majda, Commun. Math. Phys., 94, 61 (1984)] criterion, based on the blowup of vorticity, which applies to singularities of Navier-Stokes equations. We illustrate our discussion using tomographic PIV data obtained inside a high Reynolds number flow generated inside the boundary layer of a wind tunnel. In such a case, BKM and DR criteria are well correlated with each other.

 \end{abstract}

\maketitle

\section{Introduction}

Viscous incompressible fluids are described by the incompressible Navier-Stokes Equations (INSE):

\begin{numcases}
\strut
\label{NSeq}
\partial_t u_i + u_j \partial_j u_i = -\frac{1}{\rho}\partial_i P + \nu \partial_j \partial_j u_i+f_i\\
\label{incomp}
\partial_j u_j =0,
\end{numcases}

where  $u_i$  is the D-dimensional velocity field, $P$ the pressure,  $\rho$  the density, $f_i$  a D-dimensional forcing and  $\nu$   the molecular viscosity. A natural control parameter for INSE is the Reynolds number $Re=LU/\nu$, which measures the relative importance of nonlinear effects compared to the viscous ones, and is built using a characteristic length $L$ and velocity $U$.  The INSE are the corner stone of many physical or engineering sciences, such as astrophysics, geophysics, aeronautics and are routinely used in numerical simulations. However, from a mathematical point of view, it is not known whether well-defined  solutions of INSE exist. For  D = 2, the existence and smoothness theorems for solutions to the INSE have been known for a long time \cite{Ladyzhenskaya68}.  For D=3 however, it is still unclear whether the INSE are a well-posed problem, i.e. whether they remain regular over sufficient large time or develop small-scale singularities. This motivated their inclusion in the AMS Millennium Clay Prize list \cite{Fefferman}.

Historically, the search for singularities in INSE was initiated by Leray \cite{Leray34} who introduced the notion of weak solutions (i.e. in the sense of distribution). This notion was used by Scheffer to prove two theorems. First, he showed that the set of singular times (i.e. times when singularities might appear) has zero half-dimensional Hausdorff measure \cite{SchefferOrsay}. Then, in \cite{Scheffer76}, he proved that if singularities appear, then the set of singular points in space (i.e. points where singularities appear) has finite one-dimensional Hausdorff measure. Later, Caffarelli et al. \cite{Caffarelli82} introduced suitable weak solutions to study the singular set in spacetime and concluded that it has zero one-dimensional Hausdorff measure. This means that if singularities do exist, they cannot curve in spacetime. Later, it was conjectured \cite{Parisi85} that they are organized into a multifractal set.\

The problem of singularities in INSE was essentially ignored by physicists, until they realized that it may have important practical consequences regarding the rate of energy dissipation in a flow, and the resulting energy cascade from large to small scales. This effect was first discussed by Onsager \cite{Onsager49} in a work that remained essentially unnoticed, until it was reformulated into a more modern shape \cite{Duchon00, Eyink06}. Onsager observed that for a non-smooth flow field, the singularities induce an additional dissipation scaling like $U^3/L$, which does not vanish in the limit of infinite Reynolds number. Later, Eyink \cite{Eyink06} proved that singularities may also produce a non-zero rate of velocity circulation. Experimentally, it has been known for a long time that the energy dissipation rate in turbulent flows tends to a constant at large Reynolds numbers \cite{taylor35}. This observation is at the core of the 1941 Kolmogorov theory of turbulence \cite{K41a,K41b,K41c,K41d}, and was interpreted by Onsager as the potential signature of singularities with scaling exponent h=1/3 \cite{Onsager49}. In view of the importance of the Kolmogorov cascade picture in the turbulence phenomenology, it thus appears as an interesting challenge to detect directly or indirectly singularities  in incompressible fluids, and characterize their universality or dynamics. 

The most natural way of detecting singularities is via numerical simulations. In view of the  scarcity of these singularities, this detection method requires solving the full INSE at large Reynolds numbers, for a time long enough so that singularities might develop. This sets  two challenges: the resolution and the computing time. Three-dimensional solutions of INSE theoretically require a number of grid point scaling like $Re^{9/4}$. The actual memory limit of the largest supercomputers bounds the largest Reynolds number that can be simulated to $Re\approx 10^4$ \cite{Rosenberg15,He}. The second limit comes from the time it takes to simulate a certain amount of evolution time. Several million hours of CPU may be needed to integrate the evolution of a fluid at large Reynolds number for a few minutes of actual fluid evolution. These two constraints actually severely limit the quest for singularities and explain why there still is no final answer about numerical detection of singularities in INSE.

Part of the numerical limitations are relaxed when performing experiments with turbulent flows. Indeed, in a well-designed experiment, one can reach fairly easily large Reynolds numbers and monitor the results for times much longer than the characteristic time scale in the flow so that enough statistics is accumulated for reliable data analysis. In the past, experimental detection of singularities of INSE has been limited by the instrumentation since only global (torque), or localised in space (pitot, hot wire) or in time (slow imaging) velocity measurements were available. With the advent of modern Particle Image Velocimetry (PIV), measurements of the velocity field at several points at the same time over the decimetric to sub-millimetric size range is now available, at frequencies from 1 Hz to 10 kHz, reviving the interest in experimental detection of singularities of INSE. \\

In the present paper, we discuss several criteria that can a priori be used for indirect or direct experimental detection of singularities in turbulent incompressible flows. In section \ref{detectvelo} we review the techniques to detect singularities directly through the velocity field. In section \ref{detectDR}, we propose a new criterion to detect singularities through the energy dissipation that they produce. This work is based on \cite{Duchon00} and this criterion will therefore be called the Duchon-Robert (DR) criterion. We show that singularities can be detected from tomographic PIV (TPIV) data as well as from stereoscopic PIV (SPIV) data using this criterion. We further discuss how  the DR criterion compares with another well known criterion called the Beale-Kato-Majda (BKM) criterion. Finally, in section \ref{Eyinkdetection}, we propose a second criterion based on the observation made by Eyink \cite{Eyink06KelvinTheorem} that the existence of singularities may break down Kelvin theorem by producing a nonzero rate of velocity circulation. Our discussions are illustrated using tomographic PIV (TPIV) data obtained inside the boundary layer of a flow generated in a wind tunnel \cite{martins15}.

\section{Singularity detection through velocity field}
\label{detectvelo}

Before moving on to the presentation of several singularity detection methods, let us clarify what is meant by "singulariy". The problem addressed in \cite{Fefferman} is to know whether the pressure and velocity fields $P(\vec x,t)$ and $\vec u (\vec x,t)$ remain in the class $C^\infty(\mathbb{R}^3,[0,\infty))$ provided that the initial condition $\vec{u}^0 (\vec x)$ is itself a $C^\infty$ divergence free vector field on $\mathbb{R}^3$, with the additional condition that the global kinetic energy must remain bounded at all times. In what follows, $\vec u$ is considered regular if and only if it remains in the class $C^\infty(\mathbb{R}^3,[0,\infty))$ (although mathematicians prefer to work with Sobolev spaces). Therefore, we will call "singularities" points in spacetime where the velocity field is not regular (i.e. $\vec u$ itself or one of its gradients (at any order) blows up). As stated in \cite{Constantin07}, even though the problem of the existence of singularities in Euler and Navier-Stokes equations are deeply connected, there are also some major differences. For instance, the existence of friction in INSE requires that in order to observe a blow-up, the velocity field itself has to blow-up whereas this is not the case for Euler equations. This leads to the direct method presented below.

\subsection{Direct method}

Mathematically, if there exists a solution to the Navier-Stokes equations which blows up in a finite time $T$, then the velocity becomes unbounded close to $T$, at the location of the singularity \cite{Fefferman}. The natural idea seems therefore to track the velocity field and locate areas where the velocity becomes very large. In practise, it is very unlikely that such method is applicable to experimental detection of singularities: it would require time and space resolved measurements, localized at the place where the singularity appears. With current technology, this means zooming over a small area of the flow (typically a few mm$^2$) and wait until a singularity appears. Since singularities are supposed to be very scarce, there is little chance that one will be able to detect one. Moreover, if the velocity is indeed very high at this location, any tracer particle in the area will move very fast and leave the observation window in an arbitrarily small time. This is a problem for PIV or PTV measurements, which are based on particle tracking.

\subsection{Multifractal analysis}
\label{MultiFractAnal}

An interesting alternative is provided by multifractal analysis, which is now a classical but powerful method to detect singularities based on statistical multiscale analysis. Classical reviews of the method are provided in \cite{Muzy91,Kestener04}. With velocity field as the input, one obtain the so-called multifractal spectrum, quantifying the probability of observation of a singularity of scaling exponent $h$ through the fractal dimension of its supporting set $D(h)$. This method has been applied to experimental measurements of one velocity component at a single point at high Reynolds numbers in \cite{Muzy91}, which showed that  the data are compatible with the multifractal picture, with a most probable $h$ close to $1/3$.
Later, Kestener et al. \cite{Kestener04} extended the method to 3D signals (3 components of the velocity field), and showed on a numerical simulation that the picture provided by the 1D measurements was still valid, with the most probable $h$ even closer to $1/3$ than what was previously obtained using experimental measurements. To our knowledge, this method has never been applied to 3D experimental data. Moreover, due to the statistical nature of the analysis, no information could be obtained regarding the possible instantaneous spatial distribution of singularities, or their dynamics.

\section{Singularity detection through energy transfers}
\label{detectDR}
\subsection{The local energy balance}
\label{detectDRenergybalance}

It is possible to obtain a new criterion for the existence of singularities by monitoring the energy dissipation. This criterion uses a  local energy balance equation that has been derived by Duchon and Robert \cite{Duchon00} using Leray's weak solution formalism \cite{Leray34} and Onsager's ideas \cite{Onsager49}. For this, they consider a sequence of coarse-grained solutions of the Navier-Stokes equations at a scale $\ell$

\begin{numcases}
\strut
\label{NSeqcg}
\partial_t u^\ell_i + u^\ell_j \partial_j u^\ell_i = -\partial_j \tau^{ij} -\partial_i P^\ell+ \nu \partial_j \partial_j u^\ell_i \\
\label{incomp}
\partial_j u^\ell_j =0.
\end{numcases}

 and then take the limit $\ell\to 0$. They derive the corresponding energy balance

\begin{equation}
\partial_t E + \partial_j \left(u^j(E+P)-\nu\partial_j E \right)= -\nu\partial_j u_i\partial^j u^i -\mathscr{D},
\label{DREbalance}
\end{equation}

where  $\mathscr{D}$ is expressed in terms of velocity increments $\delta \vec u (\vec r,\vec x) \overset{def}{=} \vec u(\vec x + \vec r) - \vec u(\vec x) \equiv \delta \vec u (\vec r)$ (the dependence on $\vec x$ is kept implicit) as

\begin{equation}
\mathscr{D}(\vec u) \overset{def}{=} \lim_{\ell\rightarrow 0} \mathscr{D}_\ell (\vec u) = \lim_{\ell\rightarrow 0} \ \frac{1}{4\ell} \int_\mathcal{V} d\vec r \ (\vec\nabla G_\ell)(\vec r) \cdot \delta\vec u(\vec r) \ |\delta\vec u (\vec r)|^2,
\label{DRfieldnotGeneral}
\end{equation}

where $G$ is a smooth filtering function, non-negative, spatially localized and such that $\int d\vec r \ G(\vec r)=1$. The function $G_\ell$ is rescaled with $\ell$ as $G_\ell (\vec r) = \ell^{-3}G(\vec r/\ell)$. As discussed in e.g. \cite{Eyink08}, $\mathscr{D}_\ell(\vec u) = O(\delta u(\ell)^3 / \ell)$, where $\delta u(\ell) = sup_{\vert \vec r \vert <\ell} \ \vert\delta \vec u(\vec r)\vert$. As a consequence, we see that if the velocity field $\vec u$ is Hölder continuous with exponent $h$ locally in space (i.e. $\delta u(\ell) \sim \ell^h$), then $\mathscr{D}_\ell(\vec u) = O(\ell^{3h-1})$. Therefore, $h > 1/3$ everywhere ensures conservation of kinetic energy in the flow i.e. $\mathscr{D}(\vec u) = 0$ (the converse statement is a priori not true). This is Onsager's assumption \cite{Onsager49}. Actually, in their paper, Duchon and Robert gave a slightly weaker assumption under which $\mathscr{D}(\vec u) = 0$. It can be seen from (\ref{DREbalance}) that $\mathscr{D}(\vec u)$ is not related to viscosity and therefore accounts for the fraction of energy dissipated (or produced) due to a lack of smoothness of the velocity field. The classical "Kolmogorov-like" case corresponds to $h=1/3$. In this case, the singularities dissipate energy independently from the viscosity, and we have Onsager's anomalous dissipation \cite{Onsager49}. As noticed by Duchon and Robert, the expression of  $\mathscr{D}_\ell (\vec u) $ also coincides with the statistical mean rate of inertial energy dissipated per unit mass derived from the anisotropic version of the Kármán-Howarth equation \cite{KHM,Frischbook}. They therefore argue that it provides a local non-random form of the Kármán-Howarth-Monin equation, valid even for anisotropic, inhomogeneous flows. To confirm this, it has been checked in the case of an axisymmetric von Kármán flow that the contribution of $\mathscr{D}_\ell (\vec u)$ over the whole volume of the experiment agrees with global direct torque measurements of the injected power within 2$\%$ at large Reynolds numbers, as long as the scale $\ell$ lies in the inertial range \cite{Kuzzay15}.
  
Note finally that the expression of $\mathscr{D}_\ell (\vec u)$ is very suitable for its implementation starting from experimental PIV velocity fields: it involves only velocity increments, which are easily computed from the velocity field. Moreover, the very structure of $\mathscr{D}_\ell (\vec u) $ makes it less sensitive to noise than classical gradients, or even than the usual Kármán-Howarth relation: indeed, the derivative in scale is not applied directly to the velocity increments, but rather on the smoothing function, followed by a local angle averaging. This guarantees that no additional noise is introduced by the procedure. Even more, the noise coming from the estimate of the velocity is naturally averaged out by the angle smoothing. 

\subsection{Detection method}
\label{method}

We saw that if there are any singularities in the flow that can be characterized by a local Hölder exponent $h \leqslant 1$ via $\delta u(\ell) \sim \ell^h$ when $\ell\to 0$, then $\mathscr{D}_\ell(\vec u) = O(\ell^{3h-1})$ \cite{Eyink08}. Previous detections of potential singularities based on multifractal analysis indicate that the most probable exponent is close to $h=1/3$ \cite{Muzy91}. This corresponds to $\mathscr{D}_\ell(\vec u)$ bounded as $\ell \rightarrow 0$. On the other hand, for $h < 1/3$, $\mathscr{D}_\ell(\vec u)$ may diverge at small scales. Our detection technique therefore consists in monitoring $\mathscr{D}_\ell(\vec u)$ as a function of $\ell$, and track potential singularities with $h \leqslant 1/3$ by selecting points where $\mathscr{D}_\ell(\vec u)$ does not decrease with decreasing $\ell$. However, our detection method does not ensure that all singularities are detected since the condition $h \leqslant 1/3$ does not necessarily mean $\mathscr{D}(\vec u) \neq 0$. 

\begin{figure}
\includegraphics[width=1\textwidth]{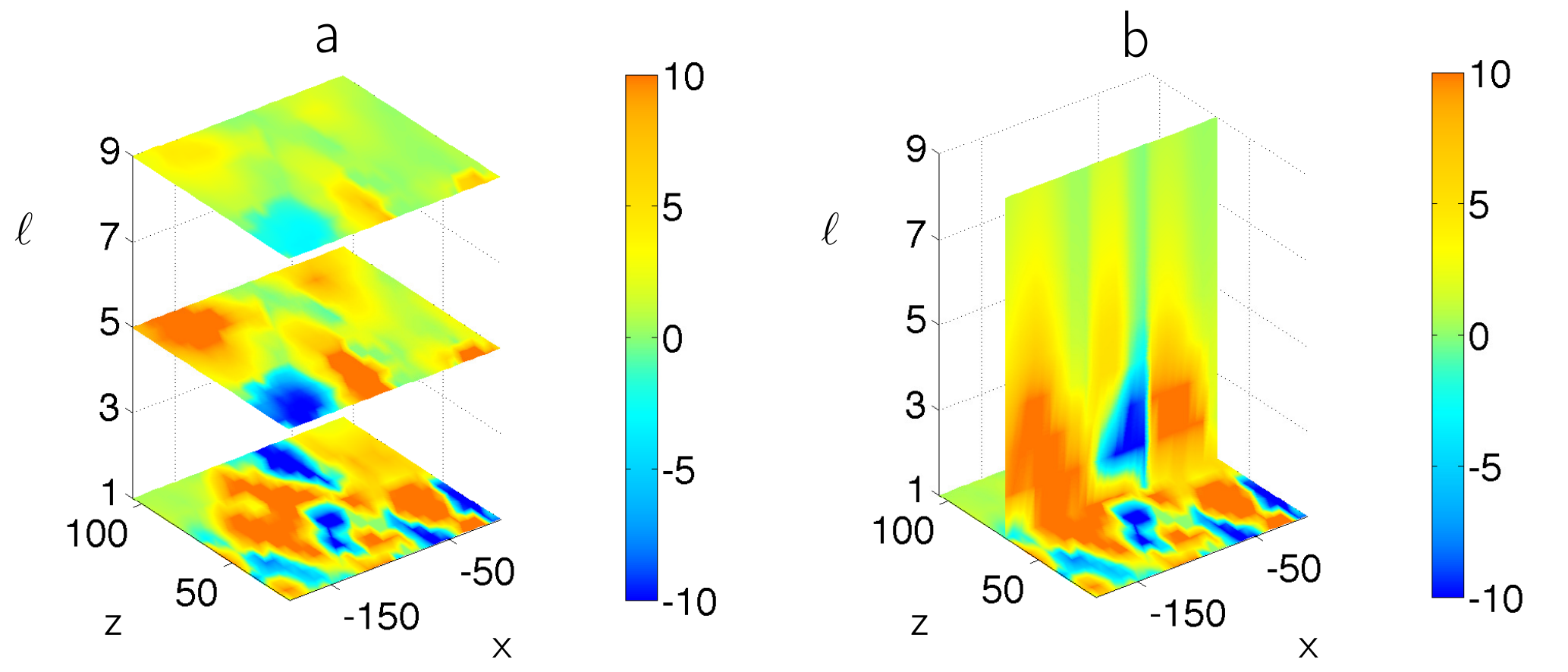}
\centering
\caption{Maps of the DR energy transfers normalized by its space-time average as a function of scale $\ell$.  a) Map of $\mathscr{D}_\ell(\vec u)$ at three different scales. b) Map of $\mathscr{D}_\ell(\vec u)$ at different scales, along a line going through a singularity. The colors code $\mathscr{D}_\ell(\vec u)$. The scale is expressed in units of the resolution scale (6 mm).}
\label{FigSacleDR}
\end{figure}

An example of variation of $\mathscr{D}_\ell (\vec u)$ as a function of scale $\ell$ and position $\vec x$ for real data is provided in Fig. \ref{FigSacleDR} (normalized by its space-time average).  For this computation, we have used a spherically symmetric function of $\vec x$ 
given by:

\begin{equation}
G_\ell(x)=\frac{1}{N}\exp(-1/(1-x^2/4\ell^2)),
\label{choixG}
\end{equation}

where $N$ is a normalization constant such that $\int d^3\vec r \ G_\ell(r)=1$. According to \cite{Duchon00}, the results should not depend on the choice of this function, in the limit $\ell\to 0$.

The experimental data are TPIV measurements performed inside a boundary layer of a wind tunnel located in the Laboratoire de Mécanique de Lille, France. The test section of the tunnel 1 m high, 2 m wide and 20 m long. The boundary layer thickness can reach up to 300 mm and the Reynolds number $R_\theta$ based on the momentum thickness is 8000. The detailed characteristics of the experimental and TPIV set-ups can be found in \cite{martins15}. We observe that as the scale $\ell$ is decreased, $\mathscr{D}_\ell (\vec u)$ does not vanish, but instead points towards localized areas which we identify as the potential singularities with $h\leqslant 1/3$. 
Note that for these data, the resolution (grid spacing) is  $\Delta x = 0.6$ mm while the Kolmogorov scale is $\eta \approx 0.35$ mm. Therefore, the flow is not fully resolved in space. In this case, detecting areas of high instantaneous energy transfers through the DR criterion is therefore a necessary but not sufficient condition to find singularities since the structures may disappear when the resolution is lowered. For this reason, when $\eta < \Delta x$, we call the corresponding  structures "potential singularities". 

\subsection{2D vs 3D detection}
\label{2Dvs3D}

In principle, this detection method requires the input of the three components of the velocity field in a volume, i.e. requires data from TPIV. In practice, many PIV systems are only stereoscopic, giving only access to the three components of the velocity field on a plane but allowing for very long statistics. Since velocity increments along one direction of space cannot be computed, this raises the question whether observing structures of DR energy dissipation from SPIV data is a sufficient condition to deduce the existence of genuine singularities inside the flow. Indeed, it could be argued that some of these structures are spurious, and would disappear if the full 3D measurements were to be performed. To answer this question, let us define a new quantity based on (\ref{DRfieldnotGeneral}), which is built from the three components of the velocity increments on a two-dimensional plane

\begin{equation}
\mathscr{D}^{2D}(\vec u) \overset{def}{=} \lim\limits_{\ell \to 0} \mathscr{D}^{2D}_\ell (\vec u) = \lim\limits_{\ell \to 0} \ \frac{1}{4\ell} \int_\mathcal{V} d\vec r \ (\vec\nabla G_\ell)(\vec r) \cdot \delta^{2D}\vec u(\vec r) \ |\delta^{2D}\vec u (\vec r)|^2,
\label{DR2D}
\end{equation}

where $\delta^{2D}\vec u(\vec r) = \vec u(\vec x^{2D} + \vec r^{2D}) - \vec u(\vec x^{2D})$, $\vec x^{2D}$ and $\vec r^{2D}$ being the projection onto the plane of measurements of the 3D coordinates. 
We now argue that areas where $\mathscr{D}^{2D} (\vec u)$ is nonzero are also areas where the full field $\mathscr{D} (\vec u)$ is nonzero, thus proving that it is sufficient to look for singularities in SPIV data. \\

To prove this, we first consider a situation where the velocity field is regular in the direction perpendicular to the plane of measurement, that we call $y$. In such a case, as $\ell\to 0$ we may  expand the velocity increments in Tayor series in the $y$-direction. Using the notations introduced in (\ref{DR2D}), we get:

\begin{equation}
\delta\vec u(\vec r) = \delta^{2D} \vec u(\vec r) + r_y \partial_y \vec u + \underset{r_y \to 0}{O}(r_y^2).
\end{equation}

where $r_y$ is the y-component of $\vec r$ and $\delta^{2D} \vec u(\vec r)$ the velocity increments on the (XZ) plane. We then take the cube of this expression which leads to

\begin{equation}
[\delta^{2D} \vec u(\vec r)]^3=[\delta\vec u(\vec r)]^3  + \underset{r_y \to 0}{O}([\delta\vec u(\vec r)]^2 r_y).
\label{deltacube}
\end{equation}

As we said in section \ref{detectDRenergybalance}, $\mathscr{D}_\ell(\vec u) = O(\delta u(\ell)^3 / \ell)$. So that if $\delta u(\ell) \sim \ell^h$, then

\begin{equation}
\mathscr{D}^{2D}_\ell(\vec u)=\mathscr{D}_\ell(\vec u) + \underset{\ell \to 0}{O}\left (\ell^{2h}\right),
\label{equafo}
\end{equation} 

where the first term is $O(\ell^{3h-1})$. So if the velocity field is regular with $h=1$, then

\begin{equation}
\lim\limits_{\ell \to 0} \mathscr{D}^{2D}_\ell(\vec u) = \lim\limits_{\ell \to 0} \mathscr{D}_\ell(\vec u)=0.
\end{equation} 

If the velocity field is singular with $h<1$, the limit of $ \mathscr{D}^{2D}_\ell(\vec u) $ is controlled by the first term of (\ref{equafo}) and so 

\begin{equation}
\lim\limits_{\ell \to 0} \mathscr{D}^{2D}_\ell(\vec u) = \lim\limits_{\ell \to 0} \mathscr{D}_\ell(\vec u).
\end{equation} 

This means that if the flow is regular in the $y$ direction, all areas where the flow is smooth in TPIV data is also smooth in SPIV data. Therefore, all singularities detected using SPIV measurements will correspond to singularities detected using TPIV. That is to say, computing the DR energy dissipation from SPIV measurements does not introduce any spurious singularities which would disappear by performing the full 3D computation. However, we cannot detect singularities lying only on the y-direction by using SPIV data. Therefore, the criterion based on $\mathscr{D}^{2D}(\vec u)$ is a sufficient detection criterion.\

\begin{figure}
\includegraphics[width=1\textwidth]{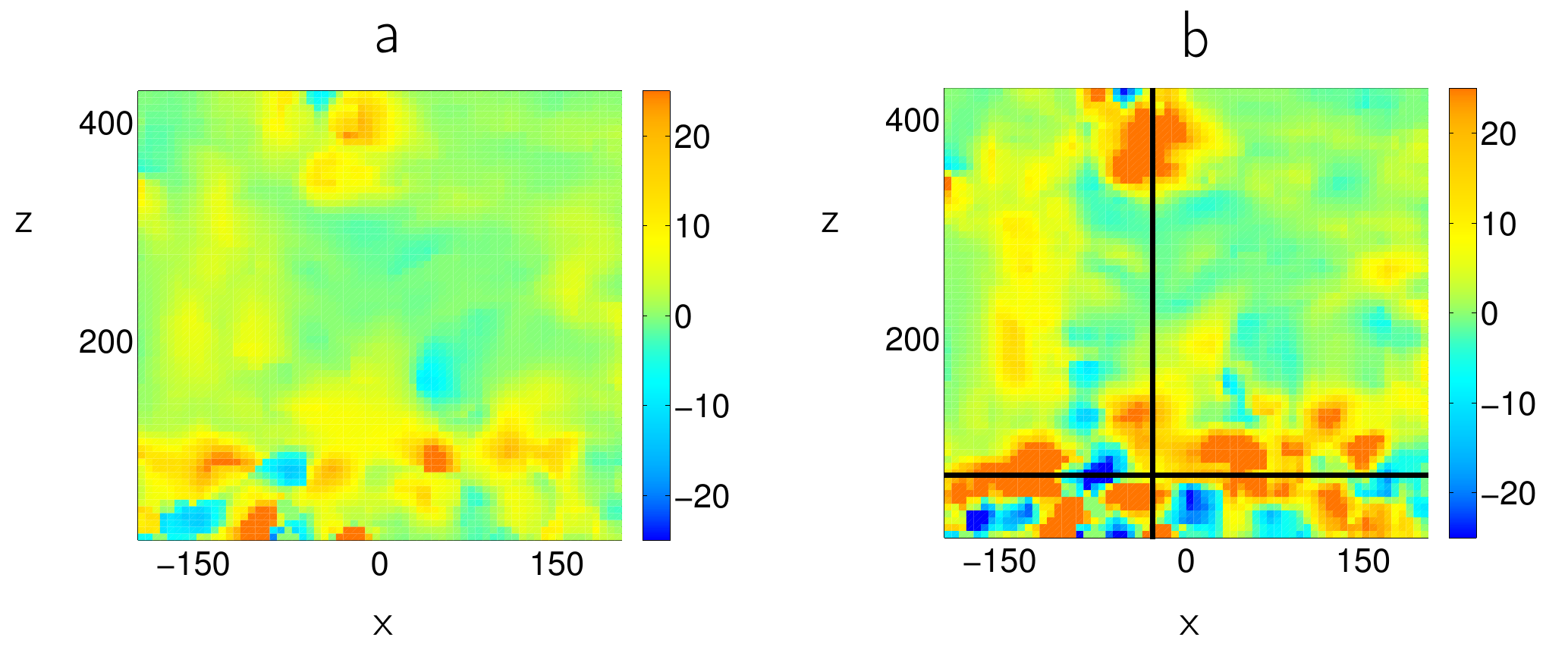}
\centering
\caption{Comparison between two instantaneous maps of a) $\mathscr{D}^{2D}_\ell (\vec u)$ and b) $\mathscr{D}_\ell (\vec u)$  normalized by their space-time averages. These two quantities have been computed from TPIV data in the boundary layer of a wind tunnel. a) and b) show the two quantities in the plane $y=0$ orthogonal to the streamwise direction. The two orthogonal lines on Fig. b) represent the two planar cuts displayed on Fig. \ref{PlanarCut}.}
\label{Comparison_2D-3D}
\end{figure}

An illustration of this can be provided by application to the boundary layer data \cite{martins15}. In such a case, there is a strong streamwise mean flow and singularities are more likely to occur in the direction orthogonal to this plane. We thus choose $y$ as the streamwise direction and compare the two criteria via instantaneous maps of $\mathscr{D}^{2D}_{\ell}(\vec u)$ (Fig. \ref{Comparison_2D-3D}a)) and $\mathscr{D}_{\ell}(\vec u)$ (Fig. \ref{Comparison_2D-3D}b)) on a plane orthogonal to the streamwise direction. Both quantities are normalized by their space-time averages.

One sees that even though there are some differences between the two maps, both fields are qualitatively the same. This confirms what we showed previously, that all areas where $\mathscr{D}^{2D}_\ell(\vec u) \ne 0$ are also areas where $\mathscr{D}_\ell(\vec u) \ne 0$. In order to quantify how much both maps are related, it is interesting to compute the Pearson's coefficient $R$ of linear correlation between areas of high energy transfer in $\mathscr{D}^{2D}_\ell (\vec u)$ and in $\mathscr{D}_\ell (\vec u)$. Here, we define "areas of high energy transfers" as being regions where the instantaneous value of the DR field is at least ten times greater than its space-time average. We find $R = 0.92$. The two fields are very well correlated, as expected. However, if one is interested in the amount of energy dissipated on the plane of interest, taking into account increments along the streamwise direction appears necessary since the space-time averages of $\mathscr{D}^{2D}_\ell (\vec u)$ is about 5 times larger than the space time-average of  $\mathscr{D}_\ell (\vec u)$ over the same plane. This may be due to contributions in the $y$ direction that have not been taken into account. Indeed,  the structures of energy dissipation appear stronger compared to their space-time average when increments along the streamwise direction are taken into account.

\begin{figure}
\includegraphics[width=1\textwidth]{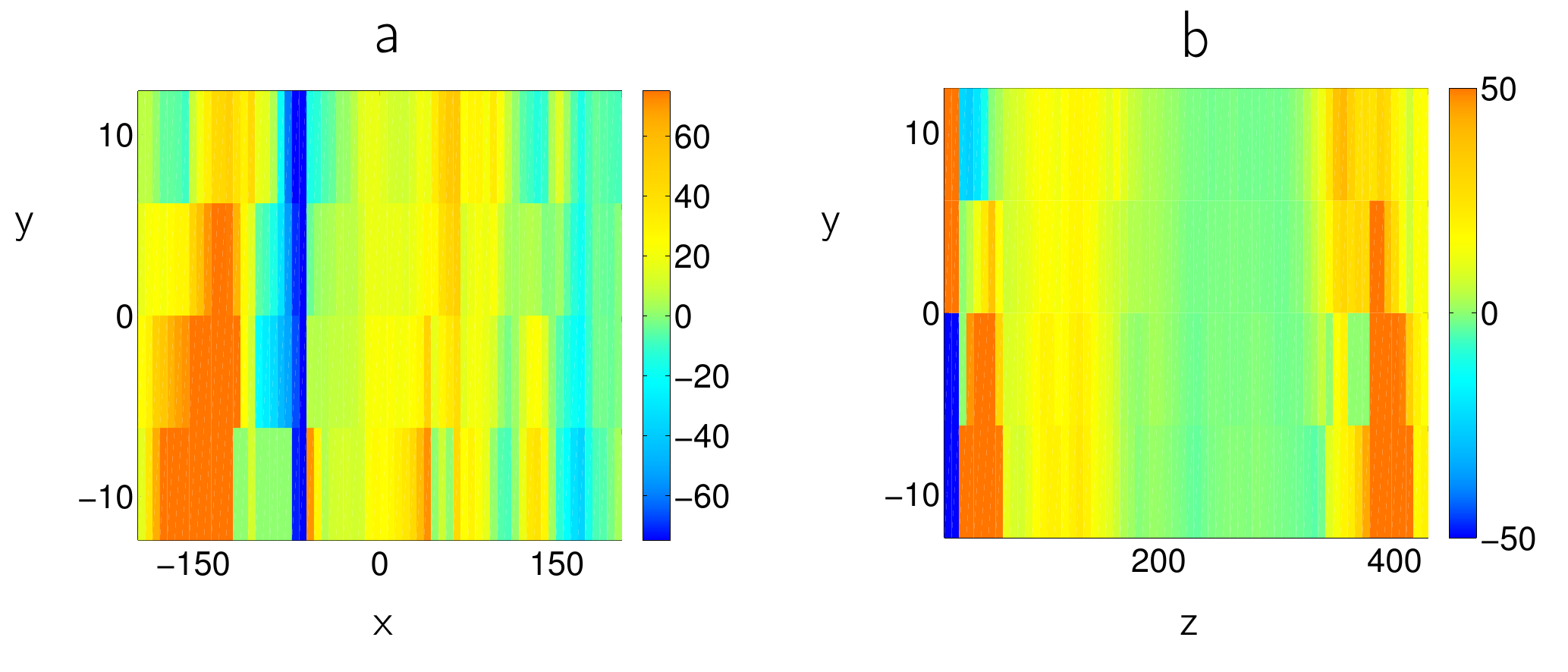}
\centering
\caption{Instantaneous maps of $\mathscr{D}_\ell (\vec u)$ normalized by its space-time average, in the two planes represented by black lines on Fig. \ref{Comparison_2D-3D}. a) shows a planar cut in an (XY) plane and b) shows a planar cut in a (ZY) plane. These maps allow us to see that singularities appear to have a three dimensional structure}
\label{PlanarCut}
\end{figure}

Fig. \ref{PlanarCut} displays two planar cuts at $z$ constant (a) and $x$ constant (b), as represented on Fig. \ref{Comparison_2D-3D}. As described in \cite{martins15}, the velocity field in only available in a few planes along the streamwise direction. Here, we have only access to five of them. Therefore, the resolution of the flow is not as good along the $y$ direction as it is for $x$ and $z$. However, we can see that singularities appear to have a three dimensional structure.

\subsection{Link with the Beale-Kato-Majda criterion}

In the limit of zero viscosity, the Navier-Stokes equations become the Euler equations. In such a case, it can be proved \cite{bkm} that if there exists a solution with a finite blowup time $T$, the vorticity $\omega(\vec x,t)$ satisfies the Beale-Kato-Majda (BKM) criterion: 

\begin{equation}
\int_0^T \vert\vert \omega(x,t)\vert\vert_\infty dt = \infty.
\label{BKM}
\end{equation}

\begin{figure}
\includegraphics[width=1\textwidth]{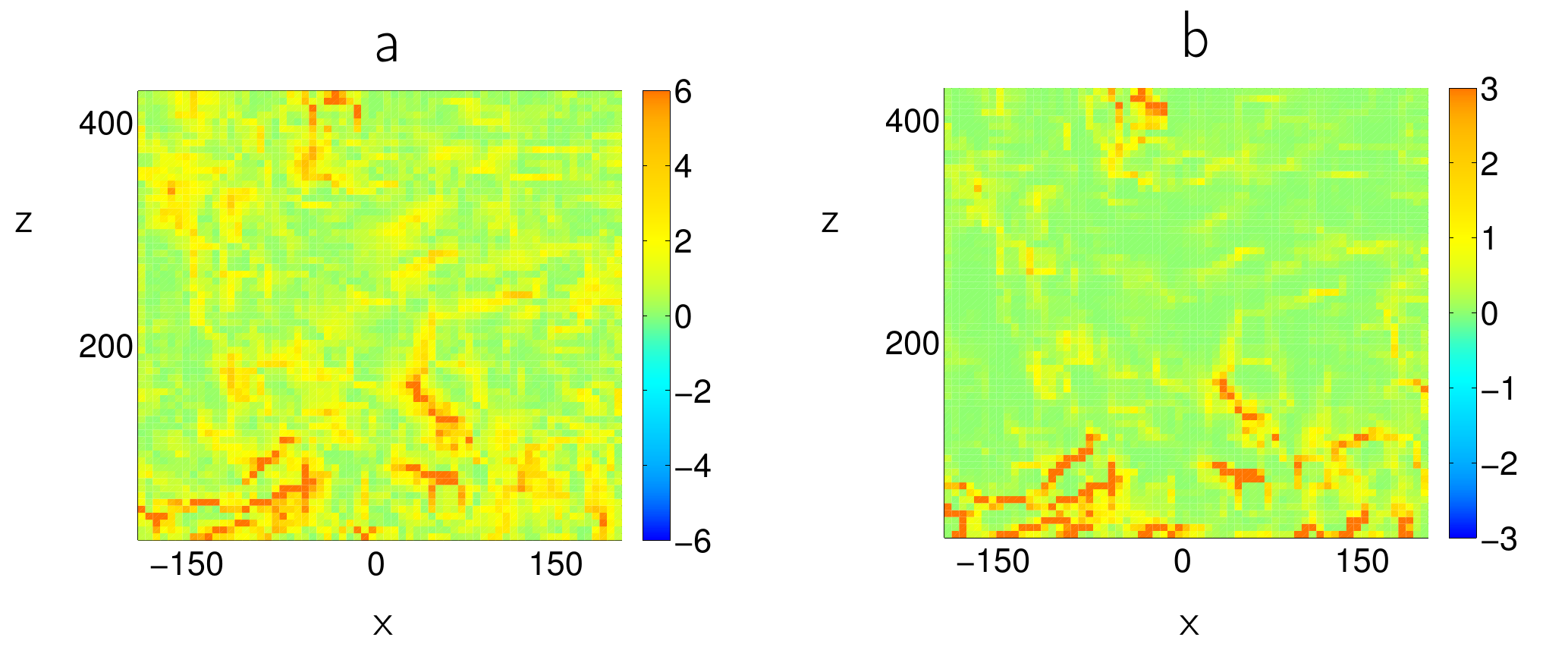}
\centering
\caption{Maps of a) $\vert\omega_y(x,z)\vert$ and b) $\vert\vec{\omega}(x,z)\vert$ computed from the same TPIV data as in Fig. \ref{Comparison_2D-3D}, and displayed in the same plane orthogonal to the streamwise direction.}
\label{vorticity}
\end{figure}

Therefore, a necessary condition for the existence of singularities is the blowup of vorticity. This criterion is usually used in numerical detection of singularities in Euler equations. However, even though the authors of \cite{bkm} only prove this result in the case of zero viscosity, they argue that their demonstration holds for nonzero viscosity, so that the theorem still applies to INSE. 
Therefore, in this section, we address the question of whether the DR and BKM criteria are correlated at large Reynolds numbers. Let us then look at Fig. \ref{vorticity}, where maps of $\vert\omega_y(x,z)\vert$ (a) and $\vert\vec{\omega}(x,z)\vert$ (b) (normalized by their space-time averages) are displayed. These quantities have been computed from the same TPIV data used in Fig. \ref{Comparison_2D-3D}, and are shown on the same planes as in Fig. \ref{Comparison_2D-3D}.

First of all, we observe on Fig. \ref{vorticity}b) that the vorticity is almost zero everywhere, except for some areas shaped as filaments where the it can be up to 60 times bigger than its space-time average. Moreover, comparing Fig. \ref{vorticity}b) with Fig. \ref{Comparison_2D-3D}b), it can be seen that areas where the structures of dissipation detected by the DR criterion are localized are also areas where the norm of the vorticity is high. As we have done between Fig. \ref{Comparison_2D-3D}a) and Fig. \ref{Comparison_2D-3D}b), we can here again perform the computation of the correlation coefficient $R_N$ between Fig. \ref{Comparison_2D-3D}b) and Fig. \ref{vorticity}b). We find $R_N = 0.59$, where the threshold of ten times the space-time average has been kept to define "areas of strong energy transfers" and "strong vorticity". Therefore, we deduce that areas of strong energy transfers in $\mathscr{D}_\ell (\vec u)$ are well correlated with areas of strong vorticity. The BKM and DR criteria are thus in good agreement.

Let us now turn to Fig. \ref{vorticity}a). This map is displayed to compare the results we obtain from TPIV data with the results we get from SPIV data. In the case of SPIV data, the only component of the vorticity that we are able to reconstruct is the orthogonal component to the plane of measurement (here $\omega_y$). Therefore, the question we ask is: does the link between the BKM and DR criteria still exists when using SPIV data? Or put another way, are areas of strong DR energy transfer also areas where $\omega_y$ is high? Comparing Fig. \ref{Comparison_2D-3D}a) with Fig. \ref{vorticity}a), there indeed seems to be a correlation between both maps. We can quantify this correlation by once again computing the correlation coefficient $R_y = 0.63$.  As a consequence, the relation between the DR and BKM criteria seems to hold well for this geometry, whether for TPIV or for SPIV.  However, there is no guarantee that it is still the same in other geometries.\\

\section{Singularity detection through circulation cascade}
\label{Eyinkdetection}

A few years after the publication of \cite{Duchon00}, Eyink noticed that singularities may also cause a breakdown of Kelvin's theorem \cite{Eyink06KelvinTheorem}, in the sense that in addition to a nonzero energy dissipation rate, they might also produce a nonzero rate of velocity circulation $\Gamma_\ell(\vec u)$ given by

\begin{equation}
\frac{d}{dt}\Gamma_\ell(\vec u) = \oint_{\mathscr{C}} d\vec s \cdot \vec{\mathscr{F}_\ell} (\vec u),
\label{circprod}
\end{equation}

where

\begin{equation}
\vec{\mathscr{F}_\ell} (\vec u) = \frac{1}{\ell} \int_\mathcal{V} d\vec r \ \left[ \left(\delta \vec u (\vec r) - \int_\mathcal{V} d\vec r' G_\ell (\vec r') \delta \vec u(\vec r')\right) \cdot \vec \nabla G_\ell (\vec r)\right] \ \delta \vec u (\vec r).
\label{DRForce}
\end{equation}

$\mathscr{C}$ being any contour advected by the fluid. $\vec{\mathscr{F}_\ell} (\vec u)$ is called the turbulent vortex-force and $G$ is a smooth filtering function having the properties given in section \ref{detectDRenergybalance}.

\subsection{Detection method}

We saw in section \ref{detectvelo} and \ref{detectDR} that the velocity field $\vec u$ of a flow might develop singularities (i.e. $\vec u$ is not $C^\infty$ anymore) due to some internal mechanisms of the INSE which are not fully understood. At the points in spacetime where this happens, $\vec u$ might however satisfy some Hölder continuity property with exponent $h$ which is a weaker regularity condition. At points where $h > 1/3$, no additional dissipation to viscosity occurs. However, if $h \leqslant 1/3$ an additional energy dissipation (or production) might appear \cite{Duchon00,Eyink08} causing kinetic energy to cascade through scales. Our detection method introduced in section \ref{detectDR} is based on the computation of this additional term to the energy balance and then track areas where it does not vanish with decreasing scale.

We introduce now a very similar detection method which is based on the observation that the turbulent vortex-force in (\ref{DRForce}) verifies $\vec{\mathscr{F}_\ell} (\vec u) = O(\delta u(\ell)^2/\ell) = O(\ell^{2h-1})$ if $\delta u(\ell) \sim \ell^h$ in the small scale limit, as discussed in \cite{Eyink06KelvinTheorem}. Therefore, the computation of the turbulent vortex-force allows us to track some but not all singularities where $h \leqslant 1/2$ whereas the DR criterion only allows us to track the ones with $h \leqslant 1/3$. Moreover, just as for the DR term (\ref{DRfieldnotGeneral}), this computation only involves velocity increments, which are easily accessible via PIV measurements. For the same reason mentioned in section \ref{method}, a detection criterion based on circulation production is only a necessary but not sufficient one (since our PIV set-up is not space resolved). Keeping the same $G$ function as in (\ref{choixG}), we can implement a detection method very similar to the one described in section \ref{detectDR}, but based on another cascading quantity. Therefore, two questions arise. Starting from our TPIV data and computing maps of $\mathscr{D}_{\ell}(\vec u)$ and $\frac{d}{dt}\Gamma_\ell(\vec u)$, are intense events in both cases well correlated? And, are we able to detect areas where a strong circulation production is observed while the DR term is weak? This would mean the detection of potential singularities with $1/3 < h \leqslant 1/2$.

\subsection{Implementation of the method} 

The arguments that have been made in section \ref{2Dvs3D} to show that it is enough to look for singularities from SPIV via energy transfers can be once again made here. Therefore, in the following, we will focus on SPIV data.

\begin{figure}
\includegraphics[width=1\textwidth]{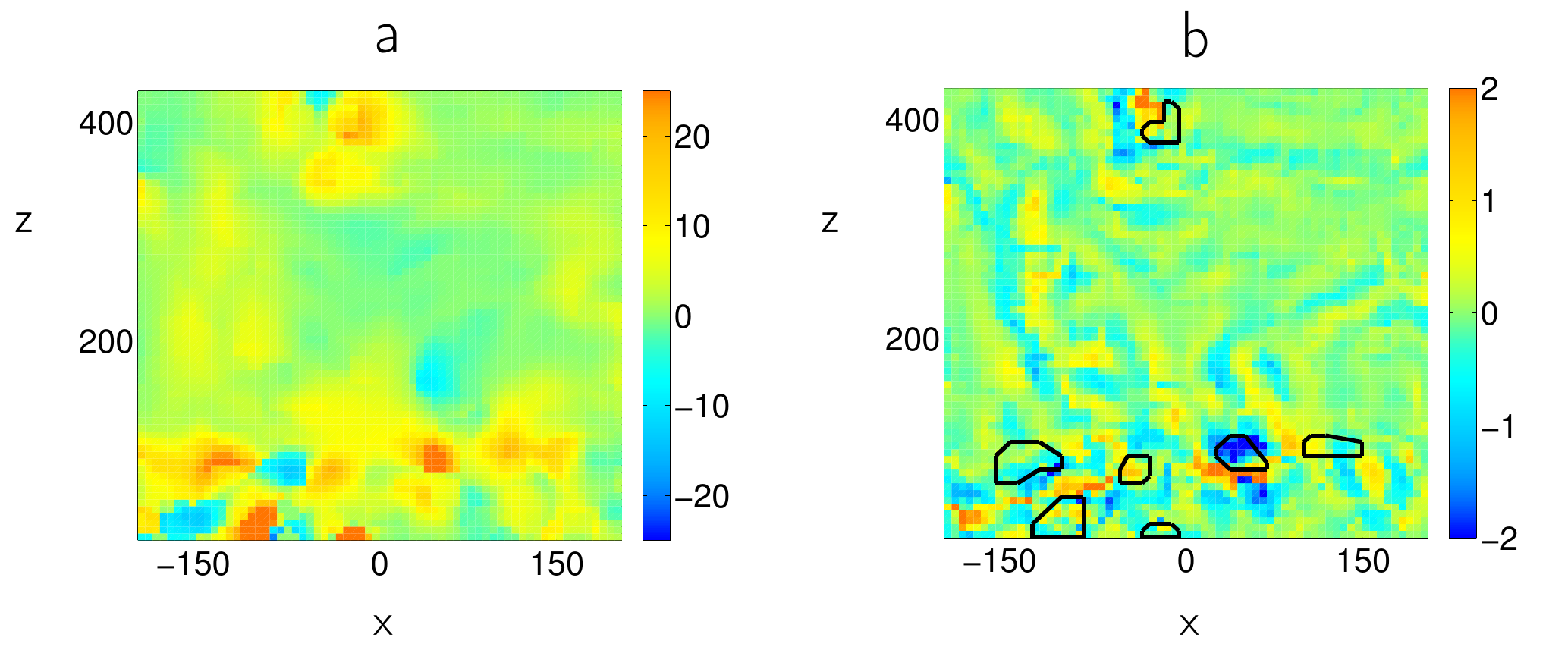}
\centering
\caption{Comparison of maps of $\mathscr{D}_{\ell}(\vec u)$ (a) and $\frac{d}{dt}\Gamma_\ell(\vec u)$ (b) for the same data set as in Fig. \ref{Comparison_2D-3D}a). For easier comparison, we have reported on the circulation map the contours of the areas where the DR dissipation is larger than 15 times its space-time average. We observe intense events in both maps. This confirms the work of Eyink that the existence of singularities might break down Kelvin theorem at very large Reynolds numbers.}
\label{Eyink-DR}
\end{figure}

Let us first compare maps of $\mathscr{D}_{\ell}(\vec u)$ and $\frac{d}{dt}\Gamma_\ell(\vec u)$ in order to answer the first question. On Fig. \ref{Eyink-DR} are displayed maps of these two quantities (normalized by their space-time averages) for the same data set as in Fig. \ref{Comparison_2D-3D}a).

First of all, it can be observed that areas where $\frac{d}{dt}\Gamma_\ell(\vec u)$ is nonzero are organized as very thin filaments. Therefore, Fig. \ref{Eyink-DR}b) is more noisy than Fig. \ref{Eyink-DR}a) even though the same procedure is applied in both cases, i.e. a derivative in scale is applied on the smoothing function, followed by a local angle averaging. There is some correlation between the maps: in areas where the dissipation is strong, there always is some nonzero circulation. However, we observe that regions of largest rate of circulation are either shifted with respect to areas of strong dissipation, or exist in some areas where there is little dissipation (see contours on Fig. \ref{Eyink-DR}b)). Overall, the Pearson's coefficient of linear correlation $R_\Gamma$ between regions of strong events in both fields is $R_\Gamma = 0.40$. This is consistent with the existence of singularities with local exponent $1/3 < h \leqslant 1/2$ that contribute mildly (or not all al) to the dissipation, but strongly to the circulation. Note that on Fig. \ref{Eyink-DR}b), areas of strong circulation production do not exceed ten times their space-time averages. Looking at other maps of $\frac{d}{dt}\Gamma_\ell(\vec u)$, it seems to be a general observation that while $\mathscr{D}_{\ell}(\vec u)$ can reach values up to 100 times its space-time average, intense events of $\frac{d}{dt}\Gamma_\ell(\vec u)$ are weaker compared to their own space-time average. Therefore, the threshold we chose here to define "intense events" has been reduced to five times the space-time average. In addition, the fact that the maps of circulation are more noisy than the maps of dissipation renders their use less straightforward to detect singularities.

\section{Discussion}

In this paper, we have introduced two new criteria based on the work of Duchon, Robert and Eyink \cite{Duchon00,Eyink06KelvinTheorem}, which allow for the local detection of singularities in experimental flows. Both criteria assume the knowledge of spatial velocity increments and are therefore easy to implement experimentally as well as numerically. The key idea behind their implementation is that velocity field in turbulent flows might satisfy Hölder continuity conditions with an exponent $h$ smaller than 1 in the limit of small scales. If $h \leqslant 1/2$, then a cascade of circulation might occur and Kelvin theorem breaks down. This cascade can be detected at larger scales provided that we are in the inertial range. In the same way, if $h \leqslant 1/3$, then a cascade of energy might occur which can also be detected in the inertial range. The first criterion that we introduced (DR criterion) focuses on these energy transfers. We showed analytically that to detect singularities, one does not need to have access to the whole velocity field inside a volume, but can instead look for them from stereoscopic particle image velocimetry (SPIV) data on a plane. This is confirmed by performing both 2D and 3D computations and comparing maps of the DR term $\mathscr{D}_{\ell}(\vec u)$. In our case, the PIV data came from the measurements of the velocity field inside the boundary layer of a wind tunnel \cite{martins15}. Clearly, being limited to SPIV data means the informations along a third direction are lacking meaning that singularities that only lies in this third direction cannot be detected. In this flow, we observe that the computation of the DR term actually shows areas where it is nonzero, some them being characterized by very strong (extreme) energy transfers through scales.

Afterwards, we compared the DR criterion to the well known Beale-Kato-Majda (BKM) criterion \cite{bkm}. We found a good agreement between them, whether SPIV or TPIV data sets are considered.

Finally, we investigated a second new criterion for the detection of singularities based on the possibility of a breakdown of Kelvin theorem at very large Reynolds numbers \cite{Eyink06KelvinTheorem}. We showed that this method is seems correlated with the DR criterion even though areas of intense energy transfers are sometimes shifted compared to areas of high rate of circulation. However, due to higher noise, this method is less reliable than the DR method, but it may allow for the detection of a wider range of singularities.

In the present paper, our detection methods were applied inside a boundary layer geometry, with data that do not reach the dissipative scale. Therefore we cannot yet conclude on the existence of singularities in experimental flows. However, the results we showed are very promising and applying our detection criteria to other types of geometries with increased resolution to check whether the structures as well as the correlations we detect still exist appears to be the next step. If this happens to be the case, this would of course not mean that the millennium problem of Navier-Stokes equations has been solved since \cite{Fefferman} requires the proof to be in $\mathbb{R}^3$ or $\mathbb{R}^3/\mathbb{Z}^3$. Indeed, experimentally we always have to generate flows inside boundaries and it is not clear what the presence or absence of these boundaries imply on the existence of singularities in incompressible Navier-Stokes equations. However, we hope our work will help providing experimental constraints on the properties of Navier-Stokes singularities as well as on corresponding suitable weak solutions. 



\bigbreak

\end{document}